\documentclass[11pt]{article}
\usepackage{jhep}
\usepackage[T1]{fontenc}
\usepackage{bm}
\usepackage{graphicx}
\usepackage{float}
\usepackage{caption}
\usepackage{subcaption}
\usepackage[normalem]{ulem}
\usepackage{gensymb}
\usepackage[utf8]{inputenc}
\usepackage{bbold}
\usepackage{soul}
\usepackage{slashed}
 \usepackage{lipsum}
\usepackage[dvipsnames]{xcolor}
\usepackage{hyperref}
\usepackage{cleveref}
\usepackage{bbold}
\usepackage{amsmath}
\hypersetup{colorlinks=false}
\usepackage{fontawesome} 
\usepackage{mathtools}
\usepackage{cancel}

\Crefname{equation}{Eq.}{Eqs.}
\Crefname{figure}{Fig.}{Figs.}

\preprint{IPPP/23/78}

\newcommand{\ogw}{\Omega_\text{gw}}

\newcommand{\equaref}[1]{Eq.~(\ref{#1})}

\newcommand{\ba}{\begin{aligned}}
\newcommand{\ea}{\end{aligned}}
\newcommand{\be}{\begin{equation}}
\newcommand{\ee}{\end{equation}}
\newcommand{\bea}{\begin{eqnarray}}
\newcommand{\eea}{\end{eqnarray}}
\newcommand{\equasref}[2]{Eqs.~(\ref{#1})~and~(\ref{#2})}

\newcommand{\figref}[1]{Fig.~\ref{#1}}

\newcommand{\secref}[1]{Section~\ref{#1}}
\newcommand{\appref}[1]{Appendix~\ref{#1}}

\newcommand{\MBH}{M_{\rm BH}}

\newcommand{\as}{a_\star}

\usepackage{mathtools}

\newcommand{\dd}{\mathrm{d}}

\def\l{\lambda}
\def\m{\mu}

\def\r{\rho}
\def\s{\sigma}

\def\del{\partial}

\title{Superradiant Leptogenesis}
\author[a]{Anish Ghoshal,}
\author[b]{Yuber F. Perez-Gonzalez,}
\author[b]{Jessica Turner}

\affiliation[a]{Institute of Theoretical Physics, Faculty of Physics, University of Warsaw, \\ ul. Pasteura 5, 02-093 Warsaw, Poland}
\affiliation[b]{Institute for Particle Physics Phenomenology, Durham University, South Road, Durham, U.K.}

\emailAdd{anish.ghoshal@fuw.edu.pl}
\emailAdd{yuber.f.perez-gonzalez@durham.ac.uk}
\emailAdd{jessica.turner@durham.ac.uk}

\abstract{
We investigate how superradiance affects the generation of baryon asymmetry in a universe with rotating primordial black holes, considering a scenario where a scalar boson is coupled to the heavy right-handed neutrinos.
We identify the regions of the parameter space where the scalar production is enhanced due to superradiance. This enhancement, coupled with the subsequent decay of the scalar into right handed neutrinos, results in the non-thermal creation of lepton asymmetry.
We show that successful leptogenesis is achieved for masses of primordial black holes in the range of order $O(0.1~{\rm g}) - O(10~{\rm g})$ and the lightest of the heavy neutrino masses, $M_N \sim O(10^{12})~{\rm GeV}$. 
Consequently, regions of the parameter space, which in the case of Schwarzchild PBHs were incompatible with viable leptogenesis, can produce the observed matter-antimatter asymmetry. 
}
\begin{document}

\maketitle
%%%%%%%%%%%%%%%%%%%%%%%%%%%%%%%%%%%%%%%%%%%%%%%%%%%%%%%%%%%%%%%%%
\section{Introduction}
\label{sec:intro}
%%%%%%%%%%%%%%%%%%%%%%%%%%%%%%%%%%%%%%%%%%%%%%%%%%%%%%%%%%%%%%%%%
The origin of the matter-antimatter asymmetry remains one of the most compelling and challenging open problems in cosmology and modern particle physics. Detailed studies of antiparticles in cosmic rays~\cite{Cohen:1997ac}, big-bang nucleosynthesis (BBN) measurements~\cite{Burles:2000ju} and very precise understanding of the cosmic microwave background radiation (CMBR) from the Planck satellite~\cite{Planck:2018vyg} provide us with stringent constraints on the number density of baryons per entropy density, $Y_B = \frac{n_b-n_{\bar{{b}}}}{s} = 8.8\times 10^{-11}$. To explain such a matter-antimatter asymmetry dynamically, a baryogenesis mechanism has to satisfy the three well-known (and necessary) Sakharov's conditions \cite{Sakharov:1967dj}: (i)  baryon number B-violation, (ii) violation of C- and CP-symmetries, and (iii) having particle interactions out of thermal equilibrium. While there are many proposed mechanisms of baryogenesis, the most well-studied are electroweak baryogenesis \cite{Kuzmin:1985mm,Shaposhnikov:1986jp,Shaposhnikov:1987tw} and baryogenesis via leptogenesis~\cite{Fukugita:1986hr} from the decays of heavy right handed (RH) neutrinos which are introduced as extension to the Standard Model (SM) to obtain light neutrino masses via Type-I seesaw mechanism \cite{see1,see2,see3,see4}.

It is well known that non-standard cosmologies affect these baryogenesis mechanisms \cite{Chen:2019etb, Abdallah:2012nm, Dutta:2018zkg, Chakraborty:2022gob, DiMarco:2022doy}. Primordial Black Holes (PBH) \cite{Carr:1974nx,Carr:1975qj} are a rich non-standard cosmology that has received renewed interest after the discovery of Gravitational Waves (GW) from black hole mergers by the LIGO-Virgo-KAGRA collaboration \cite{gw1,gw2,gw3,gw5,gw6,gw7}. With masses as tiny as $M_{\mathrm{BH}}\simeq 0.1$ g up to a few hundreds of solar masses, PBHs may lead to interesting physical signatures\footnote{Long-lived PBHs with mass ($M_{\mathrm{BH}}\gtrsim 10^{15} $g) that may survive until today may contribute to a significant energy budget of Dark Matter (DM)\cite{pdm1,pdm2,pdm3,pdm4,pdm5,pdm6,pdm7,pdm8,pdm9}}. In particular, ultralight PBHs with $M_{\mathrm{BH}}\lesssim 10^9$g, which completely evaporate before BBN ($T\sim4 $ MeV) can affect the production of dark matter \cite{Lennon:2017tqq,Gondolo:2020uqv,Cheek:2021odj,Cheek:2022dbx,Cheek:2022mmy,Sandick:2021gew} or dark radiation~\cite{Hooper:2019gtx,Lunardini:2019zob,Arbey:2021ysg,Cheek:2022dbx,Cheek:2022mmy}, the generation of
baryon asymmetry \cite{Bernal:2022pue,Perez-Gonzalez:2020vnz,Hooper:2020otu}, resonant GW spectrum induced by scalar perturbations \cite{Domenech:2021ztg}, GW from cosmic strings \cite{Ghoshal:2023sfa},  GWs induced by ultralight primordial black hole domination and evaporation \cite{Bhaumik:2022pil,Bhaumik:2022zdd} and study of vacuum stability of
the Standard Model (SM) Higgs \cite{Burda:2015isa,Burda:2016mou}. While intermediate-mass black holes $10^9{\rm g}\lesssim M_{\mathrm{BH}}\lesssim 10^{15}$g are subject to several constraints (e.g., from BBN \cite{bbn1,bbn2,bbn3}) ultralight PBHs ($0.1 {\rm g}\lesssim M_{\mathrm{BH}}\lesssim 10^9$g) face practically no cosmological or astrophysical constraints, if we assume that black holes evaporate completely, i.e., they do not leave any Planck mass-sized relics; see e.g.~\cite{pdm8, bbn1} for constraints on light PBHs coming from Planck relics.
As such, we are free to consider their interplay with new physics scenarios and how the presence of such light PBHs affects the viable parameter space of the new physics model.

PBHs may affect leptogenesis in two primary ways: (a) the production of heavy right handed neutrinos via the Hawking evaporation, which enhances the efficiency of baryogenesis via leptogenesis in some areas of the parameter space \cite{Bernal:2022pue}, and (b) entropy injection into the primordial plasma, which somewhat reduces the efficiency of leptogenesis \cite{Perez-Gonzalez:2020vnz,Calabrese:2023key,Calabrese:2023bxz}.
The black hole-induced leptogenesis scenarios, based on a type-I seesaw mechanism \cite{see4,see2,Yanagida:1979as,see1}, have been well investigated in the literature~\cite{Perez-Gonzalez:2020vnz, JyotiDas:2021shi,Bernal:2022pue,Calabrese:2023key,Schmitz:2023pfy} and it was shown that how these two effects impact leptogenesis. Generally speaking, heavy PBHs with initial masses exceeding $\gtrsim \mathcal{O}(\rm{kg})$ are typically not efficient at producing heavy right handed neutrinos with masses exceeding $10^{6}\,\rm{GeV}$, which are suitable for thermal leptogenesis but instead lead to large entropy production~\cite{Perez-Gonzalez:2020vnz,Calabrese:2023key}. Due to this, there is significant tension between such ultra-light PBH-dominated pre-BBN Universe and thermal leptogenesis. However, this picture changes when an initially spinning PBH and the presence of a scalar, which couples to the RH neutrinos, is taken into account due to the effects of superradiant instabilities. This is the subject of the present analysis.
 
More specifically, we consider a scenario where the RH neutrinos are coupled to a scalar, for instance, in a Majoron model. The final lepton asymmetry sourced from the PBHs is affected firstly due to the production of the scalar and right handed neutrinos directly from PBH evaporation via Hawking radiation, and their subsequent decays can source a lepton asymmetry. Secondly, the PBHs are very efficient at producing photons, and such entropy injections usually dilute the existing lepton asymmetry. Finally, and most importantly, since we consider rotating PBHs, some of the energy and angular momentum of the PBHs will be extracted to form bosonic clouds \cite{Penrose:1971uk,1971JETPL..14..180Z,Misner:1972kx,Starobinsky:1973aij,Brito:2015oca} of the scalar that couples to the right handed neutrinos. The subsequent decays of this scalar can produce a lepton asymmetry leading to what we call \emph{superradiant leptogenesis}. We study the interplay of these three processes and identify the regions of the parameter space that superradiance can impact the final matter-antimatter asymmetry of the Universe. For other scenarios involving superradiant production of SM and beyond-the-SM particles, see e.g.~\cite{Rosa:2017ury,Ferraz:2020zgi,Bernal:2021bbv,Bernal:2022oha,March-Russell:2022zll}

The structure of this work is as follows: in \secref{sec:SSB and Majoron oscillation}, we discuss the Majoron model we consider and follow in \secref{sec:fd} with a short review of the relevant primordial black hole physics and superradiance. In \secref{sec:BEs}, we present the Friedmann and Boltzmann equations of superradiant leptogenesis and discuss the solutions of these equations in \secref{sec:results}. Finally, we summarise and conclude in \secref{sec:disandcon}. 
In this work, we consider natural units where $\hbar = c = k_{\rm B} = 1$, and define the Planck mass to be $M_p=1/\sqrt{G}=1.22\times10^{19}\,{\rm GeV}$, with $G$ the gravitational constant.

%%%%%%%%%%%%%%%%%%%%%%%%%%%%%%%%%%%%%%%%%%%%%%%%%%%%%%%%%%%%%%%%%%%%%%%%%%%%%%%%%%%%%%%%%%%%%%%%%%%%%%%%%
\section{Spontaneous $B-L$ symmetry breaking}\label{sec:SSB and Majoron oscillation}
Throughout, we consider a  specific model that includes all the essential components for superradiant leptogenesis. It extends the SM to include right handed neutrinos and an additional scalar boson denoted as $\phi$. Notably, $\phi$ will be produced via superradiance and is also responsible for generating the RH neutrino masses. Specifically, the RH neutrino fields, denoted as $N$, acquire heavy Majorana masses through the spontaneous breaking of the global $B-L$ symmetry by $\phi$, often referred to as the Majoron. To achieve this, we introduce an SM singlet complex scalar field called $\sigma$, which carries a $B-L$ charge of $-2$. The Lagrangian is as follows:
\be\label{eq:Lagrangian}
\ba
 {\cal L} 
	& = 
	{\cal L}_{SM} 
	+ i\bar N \slashed{\del}N
	+ |\del_\m\s|^2
	- \left(Y \overline{L} \tilde{H} N
	+ \frac{g}{2} \s \overline{N^C} N + \text{h.c} \right)
	- V(H,\s)\,,\\
	V(H,\s)
	& = 
	\l_H (|H|^2 - v_{ew}^2)^2 + \l_\s (|\s|^2 - v_{B-L}^2)^2 + \l_{\s H} (|\s|^2 - v_{B-L}^2)(|H|^2 - v_{EW}^2)\,,
\ea
\ee
where  $L$ is the leptonic doublet, $H$ denotes the Higgs doublet, and $\tilde{H}=i\sigma_2H^*$. The symbols $Y$, $g$, $\l_H$,  $\l_{\s H}$, and $\lambda_{\s}$ represent dimensionless coupling constants. Additionally, we have two dimensionful parameters, $v_{EW}$ and $v_{B-L}$, which denote the electroweak and $B-L$ symmetry-breaking scales, respectively. Notably, $g$, which parameterises the strength of the scalar coupling to the RH neutrinos, will play a key role in the superradiant leptogenesis process from PBH.
Around the ${B-L}$ breaking vacuum, we parameterise $\s$ by
\begin{eqnarray}
	\sigma &=& (v_{B-L} + \r/\sqrt{2})e^{i\phi/(\sqrt{2}v_{B-L})}\,,
\end{eqnarray}
where $\rho$ is a real part of $\s$, and $\phi$ corresponds to the Nambu-Goldstone boson, i.e. the Majoron. After $B-L$ symmetry breaking, $N$ acquires the Majorana mass $M_N=g v_{B-L}$ and the Majoron acquiring a mass of order $M_{\phi} \sim \lambda_{\sigma} v_{B-L}$ in the limit the Majoron-Higgs coupling is vanishing, $ \lambda_{\sigma H} =0$. Also, this symmetry breaking leads to the formation of global cosmic strings which results in the generation of gravitational waves as we will discuss in  \secref{sec:GCS}. After electroweak symmetry breaking, the familiar type-I seesaw mechanism \cite{see1,Yanagida:1979as,see2,see4} provides mass to the light neutrinos, $m_\nu=v^2 Y M_N^{-1} Y^T$. In what follows, we will assume that the Majoron mass always exceeds twice the right handed neutrino masses, $M_{\phi}>2M_{N_{1}}$ where $M_{N_{1}}$ is the mass of the lightest RH neutrino. While at least two 
RH neutrinos are required for the successful generation of SM neutrino masses via type-I seesaw; we assume that  
$M_{\phi}< M_{N_{2}}, M_{N_{3}}$. Hence, the 
decays of $\phi$ to $N_1$  are kinematically allowed while suppressed to $N_2$ and $N_3$. The decay width of $\phi$ to RH neutrinos is
\begin{eqnarray}
\Gamma_{\phi\to N N} & = & \frac{g^2 M_{\phi}}{16\pi }\sqrt{1-\frac{4M_{N}^{2}}{M_{\phi}^{2}}}\,,
\label{eq:decaywidth_to_N}
\end{eqnarray}
where, for simplicity, we denote $N_1 \equiv N$ and $M_{N_1} \equiv M_N$ throughout the remainder of the paper.
Let us note, however, that there could be couplings between the Majoron and the other SM degrees of freedom. 
To parametrise such a possibility in a model-independent way, we define the branching ratio 
\begin{align}
    {\rm BR} \equiv \frac{\Gamma_{\phi\to {NN}}}{\Gamma_{\phi}},
\end{align}
with $\Gamma_{\phi\to {\rm NN}}$ the decay width into RH neutrinos, and $\Gamma_{\phi}$ the total $\phi$ decay width. 
Essentially, ${\rm BR}$ is dependent on the parameter $\lambda_{\sigma H}$, so increasing or decreasing $\lambda_{\sigma H}$ will increase or  decrease ${\rm BR}$.
From here onwards, we treat the only independent variables for our analysis to be $M_N$, $M_{\phi}$, $g$ and ${\rm BR}$.

%%%%%%%%%%%%%%%%%%%%%%%%%%%%%%%%%%%%%%%%%%%%%%%%%%%%%%%%%%%%%%%%%
\section{Rotating Primordial Black Holes: Hawking Radiation and Superradiance}
\label{sec:fd}
The formation PBHs during the Universe's early stages of the Universe's evolution \cite{Hawking:1975vcx, Carr:1974nx} may occur when density fluctuations exceed a certain threshold and reenter the Hubble horizon and collapse \cite{Villanueva-Domingo:2021spv}. Alternatively, PBH may form from the collapse of cosmic string loops~\cite{Honma:1991na}, domain walls, or the collision of bubbles \cite{Carr:2020xqk}. Since we are interested in superradiance studies, initially spinning PBH can be formed during aspherical collapse or in the presence of large inhomogeneities during a matter-domination era, as shown in Refs. \cite{Saito:2023fpt,DeLuca:2019buf} or from collapse of domain walls involving gravitational torques \cite{Eroshenko:2021sez} or from strong first-order phase transitions \cite{Conaci:2024tlc,Banerjee:2023qya}. We assume that an initial density of PBHs was generated during a radiation-dominated period after inflation. Considering the initial plasma temperature to be $T = T_{\rm in}$, the initial monochromatic mass of the PBH, $M_{\rm in}$, is proportional to the horizon mass \cite{Carr:2009jm,Carr:2020gox}:
\be
M_{\mathrm{in}} \equiv M_{\mathrm{BH}}\left(T_{\mathrm{in}}\right)=\frac{4 \pi}{3} \kappa \frac{\rho_R\left(T_{\mathrm{in}}\right)}{H^3\left(T_{\mathrm{in}}\right)}\,,
\ee
where $\kappa=0.2$, $\rho_R\left(T_{\mathrm{in}}\right)$  is radiation energy density at the time of PBH formation and $H$ is the Hubble rate at the time of PBH formation. The initial energy density of PBHs can then be parametrised in terms of
\be
\beta \equiv \frac{\rho_{\mathrm{BH}}\left(T_\mathrm{in}\right)}{\rho_R\left(T_\mathrm{in}\right)}=\frac{n_{\rm BH} M_{\mathrm{in}}}{\rho_R\left(T_\mathrm{in}\right)}\,,
\ee
where $\rho_{\mathrm{BH}}\left(T_\mathrm{in}\right)$ is the energy density of the PBHs at the formation time, and $n_{\rm BH}$ is the initial number density of PBHs. We assume that the population of PBHs are Kerr black holes that carry non-zero total angular momentum, $J$ \cite{Harada:2017fjm, Kuhnel:2019zbc, Eroshenko:2021sez, Flores:2021tmc, Chambers:1997ai, Taylor:1998dk, Calza:2021czr} which further is characterised by the dimensionless quantity, $a_\star = J M_p^2/\MBH^2$ where $\MBH$ the instantaneous PBH mass. 
The PBH population is assumed to have an initial monochromatic spin denoted by $\as^{\rm in}$.

The Hawking emission rate for any particle species is given by~\cite{Hawking:1974rv,Hawking:1974sw}:
\be\label{eq:GBF}
\frac{\mathrm{d}^2 \mathcal{N}_i}{\mathrm{~d} p \mathrm{~d} t}=\frac{g_i}{2 \pi} \sum_{l=s_i} \sum_{m=-l}^l \frac{\mathrm{d}^2 \mathcal{N}_{i l m}}{\mathrm{~d} p \mathrm{~d} t}\,,
\ee
with
\be
\frac{\mathrm{d}^2 \mathcal{N}_{i l m}}{\mathrm{~d} p \mathrm{~d} t}=\frac{\Gamma_{s_i}^{l m}\left(M, p, a_{\star}\right)}{e^{ \left(E_i-m \Omega\right) / T_{\mathrm{BH}}}-(-1)^{2 s_i}} \frac{p}{E_i}\,,
\ee
where individual particle species $i$ is characterised by its three-momentum $p$, total energy $E_i$, and internal degrees of freedom $g_i$, spin $s_i$.
$l$ and $m$ are the total and axial angular momentum quantum numbers, respectively, and $\Omega$ represents the angular velocity of the black hole's horizon and is given by $\Omega = (a_\star/2G\MBH)(1/(1+\sqrt{1-a_\star^2}))$. 
Finally, $\Gamma_{s_i}^{l m}\left(M, p, a_{\star}\right)$ in the Hawking spectrum represents the absorption probability of the $l,m$ mode.
The evolution of the spin and mass of a PBH without including superradiant effects is governed by the following coupled equations~\cite{Page:1976df, Page:1976ki,MacGibbon:1991tj,Cheek:2021odj}
\begin{align}\label{eq:BHmassspin}
        \frac{d\MBH}{dt} &= -\varepsilon\left(\MBH, a_\star\right) \frac{M_p^{4}}{\MBH^{2}}\,, \\
        \frac{d a_\star}{d t} &=-a_\star\left[\gamma\left(\MBH, a_\star\right)-2 \varepsilon\left(\MBH, a_\star\right)\right] \frac{M_p^{4}}{\MBH^{3}}\,,
\end{align}
where $\varepsilon\left(\MBH, a_\star\right)$ and $\gamma\left(\MBH, a_\star\right)$ are dimensionless evaporation functions related to the power and torque emission, respectively, and are obtained numerically via, see Refs.~\cite{Cheek:2021odj, Cheek:2022dbx}
\begin{align}
\varepsilon(\MBH, a_\star) &= \sum_{i}\frac{g_i}{2\pi}\int_{0}^\infty \sum_{l=s_i}^\infty\sum_{m=-l}^l E \frac{\dd^2 \mathcal{N}_{ilm}}{\dd p\dd t}\, dE\,,\\
 \gamma(\MBH, a_\star) &= \sum_{i}\frac{g_i}{2\pi}\int_{0}^\infty \sum_{l=s_i}^\infty\sum_{m=-l}^l m \frac{\dd^2 \mathcal{N}_{ilm}}{\dd p\dd t}\, dE\,.
\end{align}
In addition to Hawking evaporation being an important source of particle production, superradiance will play a key role in our scenario of interest. 

Superradiance is an enhancement phenomenon that occurs for rotating black holes, among other physical systems~\cite{Cardoso:2005vk}.
In a nutshell, a superradiant instability occurs when the occupation number of massive bosonic fields gravitationally bound to a black hole grows exponentially ~\cite{Penrose:1971uk,1971JETPL..14..180Z,Misner:1972kx,Starobinsky:1973aij,Cardoso:2005vk}.
Some conditions are required for such a phenomenon to take place.
First, the Compton wavelength of a massive bosonic field should be comparable to the BH gravitational radius in order for a bound state to be formed.
This can be parametrised via the gravitational fine-structure constant, which indicates the strength of the gravitational binding of such a state \cite{Cardoso:2005vk,Arvanitaki:2014wva}
\begin{equation}
  \alpha = \frac{M_{\phi} M_{\mathrm{BH}}}{M_p^2}\approx 0.38\left(\frac{\MBH}{10~{\rm g}}\right)\left(\frac{M_{\phi}}{10^{13}~{\rm GeV}}\right),
\end{equation}
which naturally scales with the mass of the bosonic field and black hole. 

The behaviour of the bounded state can be understood by solving the equation of motion of the gravitationally coupled scalar field in the Kerr spacetime, see e.g.~Ref.~\cite{Dolan:2007mj}. 
The energies of these \emph{quasinormal} states will be complex, $\omega =\omega_R + i\omega_I$, where the imaginary part indicates whether the bound states decays, $\omega_I<0$, or if the occupation number grows exponentially, $\omega_I>0$~\cite{Damour:1976kh, Zouros:1979iw, Detweiler:1980uk, Furuhashi:2004jk, Cardoso:2005vk, Dolan:2007mj, Rosa:2009ei, Rosa:2012uz, Dolan:2012yt, Brito:2015oca, East:2017ovw, East:2017mrj, Dolan:2018dqv}. 
Once the conditions for the formation of bound states are met, a second requirement for superradiant instability to occur is that the real part of the energy $\omega_R$ must be less than a critical value $\omega_c$, $\omega_R < \omega_c$, with
\begin{align}
    \omega_c = m\Omega\,,
\end{align}
where $m$ is the azimuthal quantum number. 
Defining $\Gamma_{\mathrm{sr}}  = \omega_I$ as the decay or growth rate, we can determine the time evolution of the number of particles $\mathcal{N}_\phi$ in the bound state through
\begin{align}\label{dNdt}
    \frac{\dd\mathcal{N}_\phi}{\dd t} = \Gamma_{\mathrm{sr}}\, \mathcal{N}_\phi\,,
\end{align}
where we have neglected, at this moment, the possibility that the scalar field $\phi$ could decay into other particles.
Moreover, note that if the particles $\phi$ have significant self-interactions, the time evolution of the cloud will be significantly modified, see Ref.~\cite{Baryakhtar:2020gao}.
For simplicity, we neglect the effect of $\phi$ self-interactions.
The generation of these particles depletes the BH mass and spin. 
Using energy conservation arguments, we have the system of equations for $\MBH$ and $a_\star$,
\begin{subequations}\label{eq:BHmassspin_SR}
\begin{align}
    \frac{\dd \MBH}{\dd t} &= - M_\phi \Gamma_{\mathrm{sr}}\, \mathcal{N}_\phi\,,\\
    \frac{\dd a_\star}{\dd t} &= -(\sqrt{2} - 2\alpha a_\star) \Gamma_{\mathrm{sr}}\frac{\mathcal{N}_\phi }{G^2 \MBH^2}\,.
\end{align}
\end{subequations}
Although there exist analytical approximations to determine the growth rate $\Gamma_{\mathrm{sr}}$ for different limits, we will apply a numerical method that determines such a rate for all values of $\alpha$ and the spin parameter $a_\star$.
For such a purpose, we follow the procedure established in Ref.~\cite{Dolan:2007mj} to solve the Klein-Gordon equation in the Kerr background.
To summarise, the procedure is as follows.
Using an ansatz in the form of a power series to solve the equations obtained after a variable separation, one obtains a system of algebraic equations for the bound state frequencies.
Employing a root finder for such an algebraic system, we thus obtain the frequencies $\omega$ that yield a bound state for a given set of quantum numbers $l,m$.
Finally, the growth rate is obtained by taking the imaginary part of the bound state frequencies.
As a cross-check, we have compared our results for the growth rate of instabilities with those in Ref.~\cite{Dolan:2007mj} and found a good agreement.

%%%%%%%%%%%%%%%%%%%%%%%%%%%%%%%%%%%%%%%%%%%%%%%%%%%%%%%%%%%%%%%%%%%%%%%%%%%%%%%%%%%%%%%%%%%%%%%%%%%%%%%%%%%%
\section{Boltzmann Equations for Superradiant Leptogenesis}\label{sec:BEs}
For the PBH mass and spin values that we investigate, the scalar $\phi$ will be sourced by both Hawking evaporation and superradiance. 
In the case that such a scalar is unstable and decays into right handed neutrinos, we will have an additional source of non-thermal leptogenesis.
In this section, we investigate the interplay between the two mechanisms. 
The set of coupled Boltzmann equations that governs the evolution of this system is
\begin{equation}\label{eq:coupled}
\begin{aligned} 
a H\frac{d n_{\phi}^{\mathrm{sr}}}{d a} & =\Gamma_{\mathrm{sr}} n_{\phi}^{\mathrm{sr}} - \Gamma_{\phi\to NN}n^{\mathrm{sr}}_{\phi} \\
aH\frac{d M_{\mathrm{BH}}}{da} & =-\varepsilon \frac{M_p^4}{M_{\mathrm{BH}}^2}-M_{\phi} \Gamma_{\mathrm{sr}} n_{\phi}^{\mathrm{sr}} \\
aH\frac{d a_{\star}}{d a} & =-a_{\star}[\gamma-2 \varepsilon] \frac{M_p^4}{M_{\mathrm{BH}}^3}-\left[\sqrt{2}-2 \alpha a_{\star}\right] \Gamma_{\mathrm{sr}} n_{\phi}^{\mathrm{sr}} \frac{M_p^2}{M_{\mathrm{BH}}^2} \\
 a H \frac{d n_{N}^{\mathrm{TH}}}{d a} & =-\left(n_{N}^{\mathrm{TH}}-n_{N}^{\mathrm{eq}}\right) \Gamma_{N}^T \\
 a H \frac{d n_{N}^{\mathrm{BH}}}{d a} & =-n_{N}^{\mathrm{BH}} \Gamma_{N}^{\mathrm{BH}}+n_{\mathrm{BH}} \Gamma_{\mathrm{BH} \rightarrow N}-n_{N}^{\mathrm{BH}} \Gamma_{\phi}^{\mathrm{BH}}+n_{\mathrm{BH}} \Gamma_{\mathrm{BH} \rightarrow \phi \rightarrow N}\\
  a H \frac{d n^{\phi}_{N}}{d a} & =2 \Gamma_{\phi \to NN}n_{\phi}^{\mathrm{sr}} - \Gamma_{N}n^{\phi}_{N}\\
  a H \frac{dn_{B-L}}{d a} & =\epsilon\left[\left(n_{N}^{\mathrm{TH}}-n_{N}^{\mathrm{eq}}\right) \Gamma_{N}^T+n_{N}^{\mathrm{BH}} \Gamma_{N}^{\mathrm{BH}} + n^{\phi}_{N}\Gamma_N\right]- \mathcal{W}n_{B-L}\,,
\end{aligned}
\end{equation}
where $a$  is the scale factor, $ n_{\phi}^{\mathrm{sr}}$, $n_{N}^{\mathrm{TH}}$, $n_{N}^{\mathrm{BH}}$, $n^{\phi}_{N}$ and $n_{B-L}$ denote the comoving number densities of $\phi$ produced from superradiance, right handed neutrinos produced from the thermal bath, Hawking radiation, scalar decays, and $B-L$ asymmetry respectively. 
The first line describes the evolution of total $\phi$ numbers due to superradiant instability, including the possible decay of $\phi$. The first (gain) term on the right-hand side (RHS), $\Gamma_{\mathrm{sr}} n_{\phi}^{\mathrm{sr}}$, corresponds to the $\phi$ production rate per BH. The second (loss) term denotes the rate of decay of these scalars into right handed neutrinos where we defined the scalar decay width, $\Gamma_{\phi\to NN}$ in \equaref{eq:decaywidth_to_N}, and we approximate that the scalar $\phi$ decays at rest~\cite{Ferraz:2020zgi}.

The second and third lines describe the evolution of PBH mass and spin discussed in \secref{sec:fd}, where we included terms coming from evaporation and superradiance, see \equasref{eq:BHmassspin}{eq:BHmassspin_SR}. The fourth tracks the number density of the right handed neutrinos produced from the thermal plasma, $n_{N}^{\mathrm{TH}}$, where $\Gamma_{N}^T$ and $n_{N}^{\text {eq}}$ are the thermally averaged decay rate and the equilibrium abundance of the right handed neutrinos, respectively.
Likewise,  the fifth line tracks the number density of the right handed neutrinos produced from the Hawking radiation. In particular, the RHS has several gain and loss terms: the second term denotes the right handed neutrinos sourced by Hawking evaporation, while the first term provides the decay rate. We note that
$\Gamma_{N}^{\mathrm{BH}}$ is the decay width of the right handed, to leptons and Higgses, corrected by an average inverse time dilatation factor, $
\Gamma_{N}^{\mathrm{BH}}\approx{\mathcal{K}_1\left(z_{\mathrm{BH}}\right)}/{\mathcal{K}_2\left(z_{\mathrm{BH}}\right)} \Gamma^{\text{rf}}_{N}$ with $\mathcal{K}_{1,2}(z)$ are modified Bessel functions of the first and second kind, respectively, and we defined $z_{\mathrm{BH}}=M_{N} / T_{\mathrm{BH}}$.
We note, however, that for the equations' full numerical solution, we use the Hawking spectrum to compute the thermal average; see for further details Refs.~\cite{Perez-Gonzalez:2020vnz,Bernal:2022pue}.
There is an additional loss and gain term from the secondary decays of scalars produced via Hawking evaporation given by the third and fourth terms, respectively, and we detail these in \appref{sec:secondary}.

The sixth line shows the dynamic evolution of the number density of right-hand neutrinos from scalar decays, where a factor of two accounts for the scalar decays into two right handed neutrinos. 
Finally, the last line involves the evolution of ($B-L$) asymmetry generated due to $N$ decays and washout, where the latter is denoted as $\mathcal{W}$. The CP-asymmetry parameter is denoted by $\epsilon$ and is a function of the Yukawa matrix, $Y$ (see \equaref{eq:Lagrangian}). We note that the Yukawa matrix is given by the usual Casas-Ibarra parametrisation $Y= 1/v \left(U \sqrt{{m}_\nu} R^T \sqrt{M}\right)$ \cite{Casas:2001sr}. Further, we assume that the light neutrinos are normally ordered and approximated that the two lightest masses to be equal, $m_1\approx m_2$, since $\Delta m^2_{21} \ll \Delta m^2_{31}$. This approximation implies that the R-matrix is a function by two real, $z = x+iy$, where we fix $x=\pi/4$ and $y = 0.44$, which ensures we are always in the strong washout regime and that $\epsilon$ is maximised allowing for the most conservative bound to be placed.   Finally, we fix the leptonic mixing angles, mass squared splitting, and CP-violating phase at their best-fit values from global
fit data \cite{Esteban:2020cvm} and fix the Majorana phases to be CP-conserving. 
This set of coupled equations, together with the Hubble expansion rate, are numerically solved using the facilities of {\sc ULYSSES} \cite{Granelli:2020pim,Granelli:2023vcm} to explore the parameter space of superradiant leptogenesis.

%%%%%%%%%%%%%%%%%%%%%%%%%%%%%%%%%%%%%%%%%%%%%%%%%%%%%%%%%%%%%%%%%%%%%%%%%%%%%%%%%%%%%%%%%%%%%%%%%%%%%%%%%%%
\section{ Leptogenesis via superradiance}
\label{sec:results}%%%%%%%%%%%%%%%%%%%%%%%%%%%%%%%%%%%%%%%%%%
\begin{figure*}
\centering
    \includegraphics[width=\linewidth]{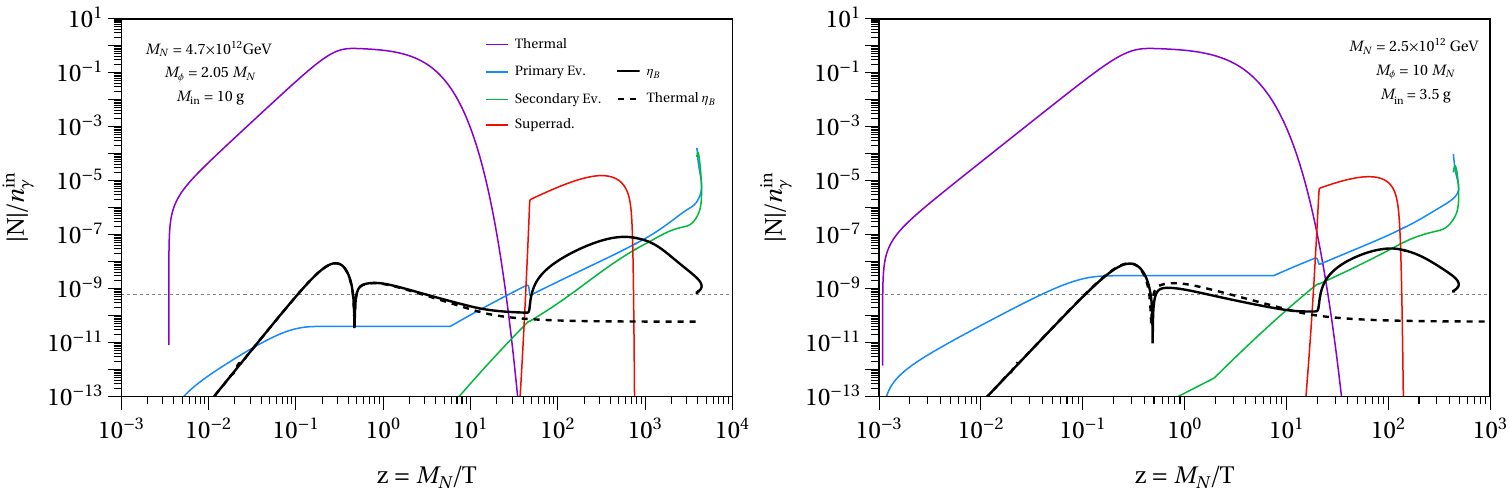}
    \caption{\label{fig:exam1} Evolution of RHN number densities normalised to the initial photon number density and the baryon-to-photon ratio $\eta_B$ for two benchmark points, $M_N=4.7\times 10^{12}~{\rm GeV}, M_\phi=2.05 M_N,  M_{\rm in} = 10~{\rm g}$ (\textbf{left}) and $M_N=2.5\times 10^{12}~{\rm GeV}, M_\phi=10 M_N, M_{\rm in} = 3.5~{\rm g}$ (\textbf{right}). We present the different species contributing to the baryon asymmetry: thermal (purple), primary from evaporation (blue), secondary from evaporation (green), and superradiance (red). The baryon-to-photon ratio is presented for the whole contribution (black) and from thermal leptogenesis only (black dashed). We assume $\as^{\rm in} = 0.999$, and $\beta = 10^{-4}$.}
\end{figure*}
%%%%%%%%%%%%%%%%%%%%%%%%%%%%%%%%%%%%%%%%%%

%%%%%%%%%%%%%%%%%%%%%%%%%%%%%%%%%%%%%%%%%%
\begin{figure}[t!]
\centering
    \includegraphics[width=\linewidth]{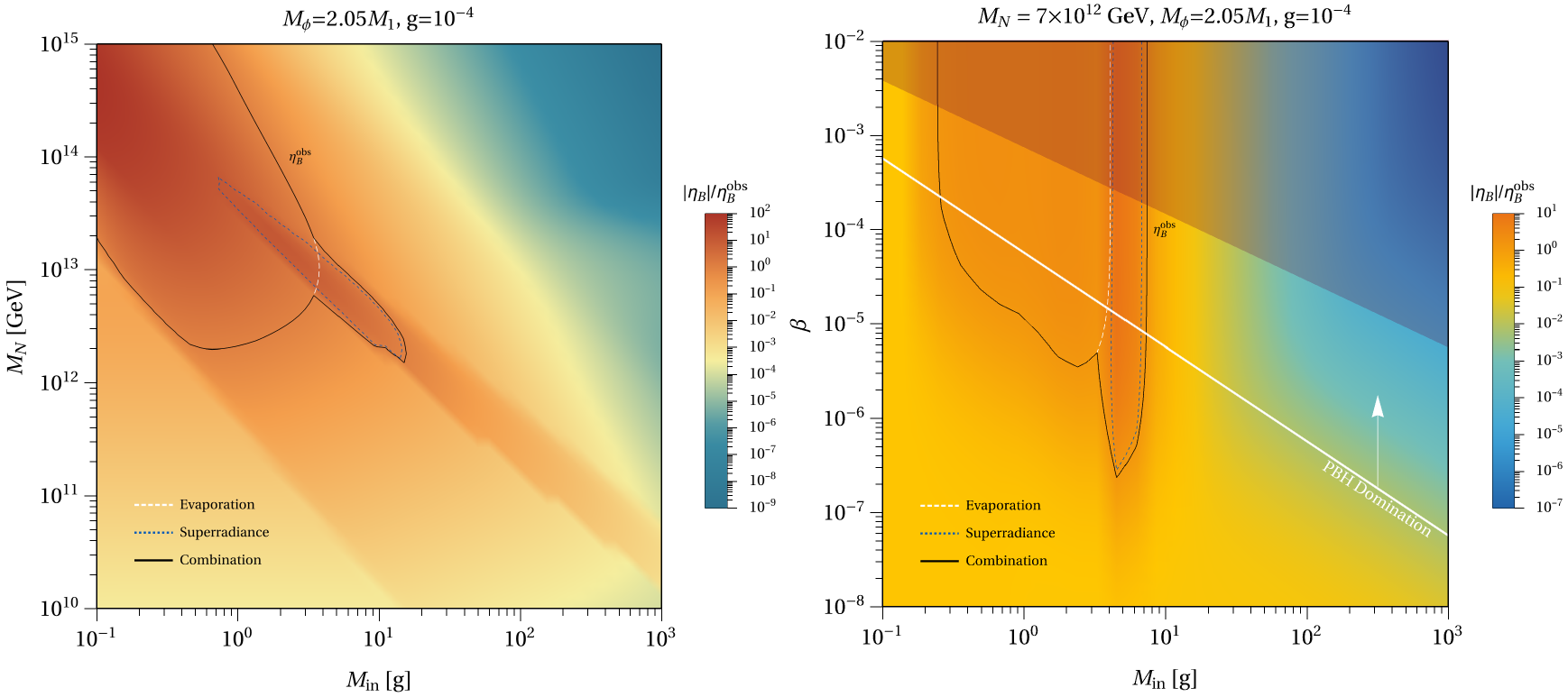}
    \caption{\label{fig:scan1} 
    \textbf{Left:} baryon-to-photon ratio normalised to the observed value in the RHN mass vs.~initial PBH mass plane. The black line indicates the parameters consistent with the observed baryon asymmetry, while the white dashed indicates the contribution from evaporation and the dotted blue from superradiance. We assume $\as^{\rm in} = 0.999$, and $\beta = 10^{-4}$. 
    \textbf{Right:} $\eta_B$ normalised to the observed value in the initial PBH fraction $\beta$ vs.~ mass plane. The black line indicates the parameters consistent with the observed baryon asymmetry, while the white dashed indicates the contribution from evaporation and the dotted blue from superradiance, assuming $M_N=7 \times 10^{12}~{\rm GeV}$. 
    The shaded region indicates the parameters excluded by GW production~\cite{Domenech:2021wkk}.
    }
\end{figure}
%%%%%%%%%%%%%%%%%%%%%%%%%%%%%%%%%%%%%%%%%%
Two example solutions for the coupled system of differential equations of \equaref{eq:coupled} are shown in \figref{fig:exam1}. In both plots, $\as^{\rm in} = 0.999$, $\beta = 10^{-4}$ while the left (right) plot the RH neutrino mass is fixed at $M_N = 4.7 \times 10^{12} $ GeV ($2.5 \times 10^{12}$ GeV) and $M_{\phi} = 2.05 \times M_N$ ($10 \times M_N$)  with initial mass of PBH  $M_{\rm in} = 10$ g (3.5 g). Further, we fix the coupling of  $\phi$ to the RH neutrinos to be $g=10^{-4}$ throughout. These two benchmark points lead to the observed baryon asymmetry once superradiance is included. The purple line shows the thermally produced RH neutrinos, while the blue and green lines show the direct production of RH neutrinos from Hawking evaporation and those produced from Hawking evaporation sourced $\phi$s that decay into RH neutrinos. The red line shows the RH neutrino population sourced from the superradiantly produced $\phi$ decays. Finally, the solid (dashed) lines represent the total (solely thermal) baryon asymmetry. We observe from both plots that the primary  RH neutrino production via Hawking evaporation follows the typical scaling, with a rapid increase in the production at the end of the PBH evaporation. The scenario with the lighter PBH (right) shows a more efficient production of RH neutrinos since the PBH temperature is higher. The secondary production occurs at later times since this relies on the initial production of the scalar $\phi$.
The production of RH neutrinos from superradiantly produced $\phi$ decays occurs at higher $z$ values for the scenario with heavier $\phi$s (right plot) because a heavier $\phi$ leads to a more significant superradiant instability growth rate and hence the bosonic cloud forms, and decays, earlier. Moreover, since the right plot shows a scenario where the scalar is an order of magnitude heavier than the RH neutrino (as opposed to a factor of a few, as shown on the left plot), the production time for RH neutrinos from this superradiance is shorter due to the increase in the $\phi$ decay width. 
As the superradiant instability exponentially amplifies $\phi$ production and its occupation, the baryon-to-photon ratio (shown in solid black)  increases by approximately two orders of magnitude. However, this increase is diminished by the large entropy injections towards the end of the PBHs lifetime. Therefore, superradiant leptogenesis can increase the efficiency of leptogenesis in some areas of the parameter space; however, the entropy dilution is always present, and we find this increase in available parameter space is rather modest.

This increase in the parameter space is presented in \figref{fig:scan1} 
where the left (right) plot shows the ratio of the predicted over the observed baryon-to-photon ratio normalised to the observed is shown in the RH neutrino mass ($\beta$) vs initial PBH mass plane. The black line shows regions that are consistent with the observed baryon asymmetry for PBH with $ a_{\rm ini} = 0.999, \beta = 10^{-4},  $ GeV and $M_{\phi} = 2.05 M_N$. Each contribution from Hawking evaporation and superradiance is also shown. We observe that the effect of superradiant leptogenesis on increasing the viable parameter space is localised in a 
horn-like region, indicated by the contour on the $M_N - M_{\rm ini}$ plane that satisfies $\alpha G M_{\mathrm{BH}} M_\phi \approx 0.2$ relation where superradiance is most significant. From the right plot, this horn-like region is not sensitive to $\beta$ (which parametrises the number density of the PBHs).

%%%%%%%%%%%%%%%%%%%%%%%%%%%%%%%%%%%%%%%%%%
\begin{figure}
\centering
    \includegraphics[width=\linewidth]{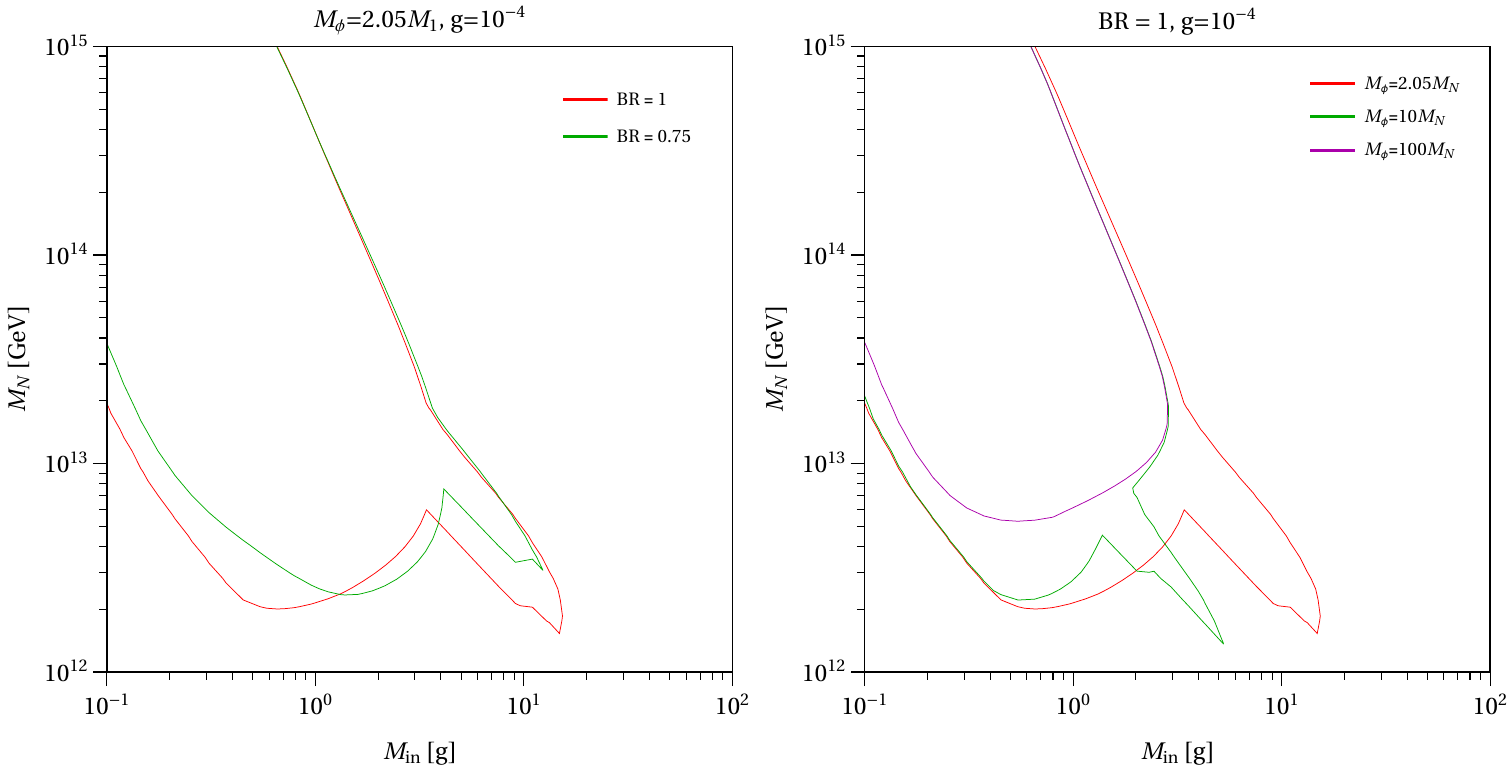}
    \caption{\label{fig:compars} Dependence of the regions consistent with the observed baryon asymmetry on the branching ratio of the scalar into RHNs (\textbf{left}) and the mass ratio between the scalar $\phi$ and the RHN mass $M_N$ (\textbf{right}). We assume $\as^{\rm in} = 0.999$, and $\beta = 10^{-4}$.}
\end{figure}
%%%%%%%%%%%%%%%%%%%%%%%%%%%%%%%%%%%%%%%%%%

The left plot of \figref{fig:compars} shows the effect of varying the $\phi$ branching ratio on the viable parameter space (red and green show the observed baryon asymmetry produced for branching ratios of $1$ and $0.75$ respectively). As expected, the parameter space shrinks for smaller branching ratios. Moreover, the effect of increasing the mass difference between the Majoron and the RH neutrino is shown on the right plot of \figref{fig:compars}. Naturally, increasing the mass of the Majoron, for a fixed RHN mass, increases the decay width and this leads to the more rapid decay of the superradiantly-produced Majoron cloud and subsequently diminishes the viable parameter space.

So far, we have exclusively examined monochromatic distributions in both PBH mass and angular momentum. The potential implications of more realistic distributions within superradiant leptogenesis is worth considering. Superradiance exhibits greater efficiency when $\alpha \sim 0.3 $. Consequently, introducing extended angular momentum distributions into our analysis is expected to reduce the contribution to leptogenesis arising from superradiant clouds.
When we introduce extended spin distributions, we anticipate a notable increase in the number of PBHs with spins below the nearly maximal value $\alpha_\star^{\rm in} = 0.999$ assumed previously. Consequently, fewer PBHs will satisfy the necessary conditions to form $\phi$ superradiant clouds, diminishing the overall effect we discussed earlier.
However, if we consider extended mass distributions, this could potentially mitigate such a reduction. This is because we might have PBHs with masses greater than $M_{\rm in}$, which could compensate for the smaller PBH spin values while maintaining $\alpha$ constant.
We nevertheless refrain from analysing this scenario quantitatively, leaving it for future work.

%%%%%%%%%%%%%%%%%%%%%%%%%%%%%%%%%%%%%%%%%%%%%%%%%%%%%%%%%%%%%%%%%%%%%%%%%%%%%%%%%%%%%%%%%%%%%%%%%%%%%%%%%%%%%%%%%%%%%%%%%%%%%%%%%%%%
\section{Primordial Gravitational Waves probes of PBH leptogenesis}\label{sec:GWs}
%%%%%%%%%%%%%%%%%%%%%%%%%%%%%%%%%%%%%%%%%%%%%%%%%%%%%%%%%%%%%%%%%%%%%%%%%%%%%%%%%%%%%%%%%%%%%%%%%%%%%%%%%%%%%%%%%%%%%%%%%%%%%%%%%%%%
In this section, we explore the various ways PBHs contribute to generating primordial gravitational waves, potentially leaving discernible imprints for specific regions of the parameter space. Firstly, it is worth noting that the formation of PBHs, characterised by significant curvature perturbations, can induce gravitational waves, especially if these perturbations are of inflationary origin \cite{Saito:2008jc} or from spectator fields present during inflation \cite{Stamou:2023vft,Chen:2023lou}\footnote{Certain scenarios may also explain the GW signal observed by NANOGrav \cite{Choudhury:2023kam,Choudhury:2023fwk}.}. Secondly, PBHs are known to emit high-frequency gravitational waves through Hawking evaporation, which serves as an additional source of such waves \cite{Anantua:2008am}. Thirdly, the merger of PBHs results in the emission of gravitational waves, thus presenting another avenue for GW production \cite{Zagorac:2019ekv}.
Fourthly, the evaporation of PBH may lead to unique GW spectral shapes emitted via cosmic strings if present in the early universe \cite{Ghoshal:2023sfa,Borah:2022iym}. Finally, the presence of inhomogeneities arising due to the distribution of PBHs also leads to density fluctuations that can induce GWs in the form of isocurvature-induced tensor perturbations \cite{Papanikolaou:2020qtd, Domenech:2020ssp, Inomata:2020lmk} as well as resonant GW signals arising due to first-order adiabatic scalar perturbation inducing second-order gravitational waves due to PBH evaporation \cite{Bhaumik:2022pil,Bhaumik:2022zdd}. Each of these gravitational wave spectra have distinct characteristics, and we will focus our discussion on the two primary sources most relevant to our specific parameter space of interest.

\subsection{Induced tensor perturbations}
Once PBHs are formed, their distribution in space is random and act as sources of inhomogeneities present during inflation in the form of isocurvature perturbations \cite{Papanikolaou:2020qtd}. If present in the early Universe, these inhomogeneities lead to density fluctuations in the form of isocurvature perturbations. Later on, when PBHs dominate the Universe's energy density, the isocurvature perturbations are converted to adiabatic perturbations that source tensor perturbations at the second order. These tensor perturbations propagating as GWs are probing the time of PBH formation. Next, the presence of adiabatic scalar perturbation (either of inflationary origin or any other source) also induces tensor perturbations and since these scalar modes are enhanced due to the almost instantaneous evaporation of PBHs effectively changing the background from matter-domination to radiation domination. This leads to unique resonant GW enhance signals that may be detected. Combining these two effects leads to a unique double-peaked GW spectrum with the present-day dimensionless energy density given by~\cite{Bhaumik:2022pil,Bhaumik:2022zdd,Domenech:2020ssp,Barman:2022pdo}
\begin{equation}
    \ogw(t_0,f)\simeq \ogw^{\rm peak}\left(\frac{f}{f^{\rm peak}}\right)^{11/3}\Theta
\left(f^{\rm peak}-f\right),\label{eqn:omgw}
\end{equation}
where  the peak of the spectrum is determined by the number density and initial mass of the PBHs
\begin{equation}
    \ogw^{\rm peak}\simeq 2\times 10^{-6} \left(\frac{\beta}{10^{-8}}\right)^{16/3}\left(\frac{M_{\mathrm{in}}}{10^7 \rm g}\right)^{34/9}.\label{eqn:omgpeak}
\end{equation}
These analyses are valid only for distances larger than the mean separation between PBHs, which in turn imposes an ultraviolet cutoff to the GW spectrum, with $f_{\text{peak}}$ corresponding to the co-moving scale representing the mean separation of PBHs at the time of PBH formation. Therefore appears the $\Theta$-function in Eq.~\eqref{eqn:omgw} with 
\begin{equation}
    f^{\rm peak}\simeq 1.7\times 10^3\,{\rm Hz}\,\left(\frac{M_{\mathrm{in}}}{10^4 \rm g}\right)^{-5/6}.\label{eqn:fpk}
\end{equation}

\subsection{Gravitational Waves from global cosmic strings}\label{sec:GCS}
The spontaneous breaking of the global $U(1)_{B-L}$ symmetry, as discussed in \secref{sec:SSB and Majoron oscillation}, can lead to the generation of cosmic strings. After this symmetry breaking, a network of horizon-size long strings can form whose length is characterised by a correlation length $L=\sqrt{\mu/\rho_\infty}$ of loop strings, where $\rho_\infty$ is the long-string energy density and $\mu $ is the string tension. These strings can intersect and form loops that can oscillate and generate gravitational waves whose energy density depends on the string tension.

The cosmic string network can reach a scaling solution, where the string energy density comprises a constant fraction of the Universe's energy density. As seen in several simulations, the size of a radiating loop at a cosmic time $t$ is given by $l(t)=\alpha t_i-\Gamma G\mu(t-t_i)$, where $l_i=\alpha t_i$ is the initial size of the loop, $\Gamma\simeq 50$~\cite{Vilenkin:1981bx}, and $\alpha\simeq 0.1$. The energy loss from a loop can written as a set of normal-mode oscillations with frequencies $f_k=2k/l_k=a(t_0)/a(t)f$, where $k=1,2,3...\infty$. The $k$th mode GW density parameter is obtained as 
\bea
\Omega_{GW}^{(k)}(f)=\frac{2kG\mu^2 \Gamma_k}{f\rho_c}\int_{t_{osc}}^{t_0} \left[\frac{a(t)}{a(t_0)}\right]^5 n\left(t,l_k\right)dt\,,\label{gwf1}
\eea
 where $t_{osc}$ is the oscillation time, $n\left(t,l_k\right)$ is the loop number density which maybe estimated using the Velocity-dependent-One-Scale (VOS) model.  The quantity $\Gamma_j$ is given by $\Gamma_j=\frac{\Gamma j^{-\delta}}{\zeta(\delta)}$, with $\delta=4/3$ for loops containing cusps \cite{Damour:2001bk}. 

Detectable GWs from the global cosmic string loops arise from the most recent epoch of cosmic evolution. The whole GW amplitude may grow with $\mu$, but in the presence of a PBH-dominated era before the most recent radiation epoch at $T = T_\Delta$, the GW spectrum at higher frequencies are given by $\Omega_{\rm GW}^{(1)}(f\lesssim f_\Delta)\sim f^0 = \text{const}$ and  $\Omega_{\rm GW}^{(1)}(f\gtrsim f_\Delta)\sim f^{-1}$, with $f_\Delta$ the frequency of the spectral-break. This leads to unique GW spectra due to the presence of PBH-domination era and its subsequent evaporation leading to the onset of radiation era and can be tested via several upcoming GW detectors as studied in detail in Ref.~\cite{Ghoshal:2023sfa}.

In addition, global cosmic strings efficiently contribute to $N_{\rm eff}$  via Goldstone boson emission. However, the precise constraint is still debated among the various groups performing lattice simulations (see e.g. Refs \cite{Hindmarsh:2019csc,Hindmarsh:2021vih,Buschmann:2019icd,Buschmann:2021sdq} and Refs. \cite{Gorghetto:2018myk,Gorghetto:2020qws, Gorghetto:2021fsn}). To be safe from this we take the upper bound to be $v_{B-L} \lesssim 10^{15} \, {\rm GeV}$ following the most stringent among Ref.~\cite{Chang:2021afa} and  Ref.~\cite{Gorghetto:2021fsn,Dror:2021nyr} \footnote{The non-observation of B-modes in CMB provides another constraint on global strings from $H_{\rm inf}\lesssim 3\times 10^{13}~\rm GeV$ (scale of inflation) \cite{BICEP:2021xfz} via the maximum temperature of the Universe $T_{\rm max} \lesssim 4 \times 10^{15}$~GeV to $v_{B-L} \lesssim 10^{15} \, {\rm GeV}$.}. This translates into a RH neutrino mass of around $M_N = 10^{11}$ GeV for $g=10^{-4}$ (see \figref{fig:compars}). For this scenario, and PBH mass of $10^{2} - 10^{3}$g, the corresponding GW spectrum will be probed via the Einstein Telescope (ET) GW detector with four years of data-taking and the signal-to-ratio needed to claim discovery is needed to be 10 or larger as studied in details in Ref. \cite{Ghoshal:2023sfa}. However, with the same $v_{B-L}$ for larger $g$ values (that is, for larger $M_{N}$) we find that the superradiance cloud decays too quickly to have any sizeable impact on the lepton asymmetry generation. Nonetheless, it could happen that based on the uncertainties related to the lattice simulation of global cosmic strings, and the sensitivity reaches of the ET GW detector, we may also be able to probe some regions of the parameter space shown in \figref{fig:compars} via this GW spectral shape and leave this for future dedicated studies. Further, if we had considered a local $B-L$ symmetry\footnote{Also see Ref.\cite{Datta:2020bht} for a similar discussion, but without PBH spin.}, then there are clear overlaps between the parameter space shown in Fig. 6 of Ref. \cite{Ghoshal:2023sfa} and \figref{fig:compars}. However, such a theory also introduces a heavy Z$_{B-L}$ vector boson, which would also have its superradiance effects, and those signatures would require a separate dedicated study which is beyond the scope of our present analysis.

%%%%%%%%%%%%%%%%%%%%%%%%%%%%%%%%%%%%%%%%%%%%%%%%%%%%%%%%%%%%%%%%%%%%%%%%%%%%%%%%%%%%%%%%%%%%%%%%%%%%%%%%%%%%%%%%%%%%%%%%%%%%%%%%%%%%
\section{Discussion and Conclusion}\label{sec:disandcon}
We have studied the impact of the superradiant primordial black holes on the production of the baryon asymmetry of the Universe via high-scale leptogenesis. 
We investigated how a Majoron model would be impacted in the context of high-scale leptogenesis from spinning black holes. 
In particular, the spinning PBHs' evaporation is associated with particle production bursts due to superradiance and an injection of scalar Majoron $\phi$, which decays into right handed neutrinos. We found that such superradiance effects can enhance the production of a lepton asymmetry beyond the region of the parameter space in the minimal scenario, and this occurs where the superradiance condition is satisfied, $\alpha G M_{\mathrm{BH}} M_\phi \approx 0.2$, see the horn-like region of \figref{fig:scan1}.
This effectively rescues some regions of the parameter space otherwise, which did not yield viable high-scale leptogenesis as the baryogenesis mechanism \cite{Perez-Gonzalez:2020vnz}. This additional viable parameter space requires a PBH population with masses from $0.1$ to $10$ grams with an initial abundance $\beta \gtrsim 10^{-7}$ (see \figref{fig:scan1}). This impact on the parameter space is also governed by the microphysics BSM parameters involving RH neutrino masses and the coupling of the Marojon to SM, depicted in terms of BR (see \figref{fig:compars}).

As described in \secref{sec:GWs}, recent studies regarding presence of PBH-dominated epoch \cite{Inomata:2020lmk,Papanikolaou:2020qtd,Domenech:2020ssp,Domenech:2021wkk,Bhaumik:2022pil,Bhaumik:2022zdd, Papanikolaou:2021uhe, Papanikolaou:2022chm, Papanikolaou:2022hkg} have shown that inevitable scalar-induced gravitational waves resulting from adiabatic and isocurvature perturbations (model-independent) can produce distinctive resonant peaks and double-peaks in the GW spectrum, which can be used to probe the formation time and decay time of PBHse. The unique spectral shapes of those GW are key to the understanding if PBHs ever came to dominate the Universes' energy density \cite{Bhaumik:2022pil,Bhaumik:2022zdd}. If such spectra cover the regions where superradiance can increase the number density of right handed neutrinos produced by PBHs (via intermittent production of $\phi$)  one may perhaps have a complementary test of superradiant leptogenesis via primordial GW. However, such a study is beyond the scope of the present analysis. Moreover, gauging the B-L symmetry will have interesting consequences due to the presence of massive $Z_{B-L}$ boson, which could have its own superradiance effects as well as the GW signatures coming from local cosmic strings would show up as signatures in the GW spectrum that LISA would be able to detect.

\section*{Acknowledgments}
\medskip
This work has been funded by the UK Science and Technology Facilities Council (STFC) under grant ST/T001011/1. 
This project has received funding/support from the European Union's Horizon 2020 research and innovation programme under the Marie Sklodowska-Curie grant agreement No 860881-HIDDeN.
This work has made use of the Hamilton HPC Service of Durham University. This work used the DiRAC@Durham facility managed by the Institute for Computational Cosmology on behalf of the STFC DiRAC HPC Facility
(www.dirac.ac.uk), which is part of the National eInfrastructure and funded by BEIS capital funding via
STFC capital grants ST/P002293/1, ST/R002371/1 and
ST/S002502/1, Durham University and STFC operations
grant ST/R000832/1.

\appendix

\section{Secondary Decays}\label{sec:secondary}
The PBHs produce RH neutrinos via Hawking radiation, which can decay and produce a lepton asymmetry. The rate of this is given by:
\be
\Gamma_{\mathrm{BH} \rightarrow N} = \int \frac{d^2N_N}{dE_Ndt} dE_N\,.
\ee
However, the PBHs will also produce the scalar, $\phi$, via Hawking radiation which can undergo a secondary decay
to RHNs which will decay. These secondary RHNs (denoted by superscript 'sec') will have a slightly different energy spectrum than the primaries. Their
differential number density is given by
\begin{equation}
\begin{aligned}
\frac{d^2N^{\rm{sec}}_N}{dE_{N}dt} & = \int dE_{\phi} \frac{d^2N_\phi}{dE_{\phi}dt} \frac{dN_{\phi\to NN}}{dE_N}\,,\\
\end{aligned}
\end{equation}
where the rate of the RHNs sourced from $\phi$ decays is found by integrating the above differential number density over the energy of the RHNs:
\[
\Gamma_{\mathrm{BH} \rightarrow \phi \rightarrow N} = \int \frac{d^2N^{\rm{sec}}_N}{dE_{N}dt} dE_N\,.
\]

\bibliographystyle{JHEP}
\bibliography{reference}
\end{document}